\documentclass[%
 reprint,
showpacs,preprintnumbers,
nofootinbib,
 amsmath,amssymb,
 aps,
]{revtex4-1}

\usepackage{graphicx}
\usepackage{dcolumn}
\usepackage{bm}

\begin{document}
\title{Microscopic analysis of fusion hindrance in heavy systems}

\author{Kouhei Washiyama}
 \affiliation{RIKEN Nishina Center, Wako 351-0198, Japan}
 \email{kouhei.washiyama@riken.jp}

\date{\today}

\begin{abstract}
\begin{description}
\item[Background] Heavy-ion fusion reactions involving heavy nuclei at energies around the Coulomb barrier exhibit fusion hindrance, where the probability of compound nucleus formation is strongly hindered compared with that in light- and medium-mass systems.
The origin of this fusion hindrance has not been well understood from
a microscopic point of view. 
\item[Purpose] Analyze the fusion dynamics in heavy systems by a microscopic reaction model and understand the origin of the fusion hindrance.
\item[Method] We employ the time-dependent Hartree-Fock (TDHF) theory as a microscopic 
reaction model. We extract nucleus--nucleus potential and energy dissipation
by the method combining TDHF dynamics of the entrance channel of fusion reactions with one-dimensional Newton equation including a dissipation term. 
Then, we analyze the origin of the fusion hindrance using the properties of the extracted potential and energy dissipation.
\item[Results] We obtain finite extra-push energies for heavy systems from TDHF simulations, which agree with experimental observations. 
Extracted nucleus--nucleus potentials show monotonic increase as the relative distance of two nuclei decreases, which induces the disappearance of an ordinary barrier structure of the nucleus--nucleus potential. 
This is different from those in light- and medium-mass systems and from density-constraint TDHF calculations.
Extracted friction coefficients show sizable energy dependence and universal 
value of their magnitude,
which are rather similar to those in light- and medium-mass systems.
Using these properties, we analyze the origin of the fusion hindrance and find that contribution of the increase in potential to the extra-push energy is larger than that of the accumulated dissipation energy in most systems studied in this article.
\item[Conclusions] We find that the nucleus--nucleus potentials extracted in heavy systems show a specific property, which is not observed in light- and medium-mass systems. 
By the analysis of the origin of the fusion hindrance, we conclude that, as the system becomes heavier,
the dynamical increase in nucleus--nucleus potential 
at small relative distances plays a more important role than the dissipation during the fusion reaction for understanding the origin of the fusion hindrance.
\end{description}
\end{abstract}

\maketitle


\section{Introduction}\label{sec:introduction}

Heavy-ion fusion reactions at energies around the Coulomb barrier 
have been studied for a long time and give interesting phenomena such as 
an enhancement of fusion cross sections at energies below the Coulomb barrier. 
For a better understanding of fusion dynamics, 
it is crucial to consider the couplings between the collective motions
of colliding nuclei and their internal excitations.
To deal with such couplings, coupled-channels models have been developed 
and succeeded in reproducing such an enhancement of fusion cross sections at subbarrier energies \cite{balantekin98,dasgupta98,hagino12,back14}. 
In those models, unless a colliding system is too heavy or too light,
it has been generally considered that fusion takes place 
once colliding nuclei overcome the Coulomb barrier
because of strong absorption inside the Coulomb barrier.
Then, instead of using an imaginary potential inside the Coulomb barrier
with the regular boundary condition at the origin, 
an incoming wave boundary condition \cite{dasso83} has often been employed.
The approaching phase leading to overcoming the Coulomb barrier
to make a contact of projectile and target is considered as the capture process.

The property of low-energy fusion dynamics drastically changes when the product of the charge numbers of the projectile and target nuclei, denoted by $Z_PZ_T$, becomes greater than 1600. 
In such heavy systems, it has been known that the fusion probability
is extremely hindered around the Coulomb barrier \cite{gaggeler84,sahm84,sahm85,schmidt91}, which is called fusion hindrance\footnote{Notice that this phenomenon is different from fusion hindrance observed at deep subbarrier energies in light- and medium-mass systems \cite{jiang02,jiang04,dasgupta07}.}.
One of the well known experimental observations for exhibiting this fusion hindrance is a direct comparison between the $^{40}$Ar\,+\,$^{180}$Hf ($Z_PZ_T=1296$) and $^{96}$Zr\,+\,$^{124}$Sn ($Z_PZ_T=2000$) systems leading to $^{220}$Th reported in Ref. \cite{sahm85}.
The authors of Ref. \cite{sahm85} measured evaporation-residue cross sections of these systems to deduce their fusion probabilities for central collisions, and then compared them at the Coulomb barrier energy of both reactions estimated by the Bass model \cite{bass74}.
They reported that the fusion probability of the $^{96}$Zr\,+\,$^{124}$Sn system is hindered by two orders of magnitude from that of the $^{40}$Ar\,+\,$^{180}$Hf system and from a simple barrier penetration model and that the apparent fusion barrier for the $^{96}$Zr\,+\,$^{124}$Sn system is shifted by about 26.7\,MeV from the Bass barrier. 
Before this observation, Swiatecki and coworkers had predicted that an additional kinetic energy to the Coulomb barrier, called extra-push energy, is needed for fusion to take place 
when the so-called fissility parameter of the system (its definition given in Eqs.~(\ref{eq:fissility}) and (\ref{eq:fissility1}) in Sec.~\ref{sec:result:extrapush}) is larger than its threshold value, which approximately corresponds to $Z_PZ_T>1600 - 1800$ \cite{swiatecki81,swiatecki82,bjornholm82,blocki86}. 
They considered that, after the contact of colliding nuclei, large Coulomb force might make the colliding system reseparate under the appearance of an additional potential barrier inside the contact configuration. 
This indicates the necessity of treating the second step towards fusion, i.e.,
after the capture process as the first step,
the process of shape evolution from the contact configuration of colliding nuclei to the configuration of compound nucleus formation. 
Since the projectile and target densities are significantly overlapped 
in this formation process, strong energy dissipation is expected to arise from the collective motions to the internal excitations.
Therefore, the potential energy landscape and the mechanism of energy dissipation inside the Coulomb barrier are important quantities to understand the property of the fusion mechanism in heavy systems.

For the analysis of the competition between compound nucleus formation and quasifission in such heavy systems, especially in the context of the synthesis of superheavy elements \cite{hofmann00,oganessian07,morita04,morita12},
dynamical models such as 
a diffusion model based on a master equation \cite{adamian97,diaz-torres01,diaz-torres06}
and a macroscopic fluctuation--dissipation model based on a Langevin equation \cite{abe02, aritomo04, zagrebaev05, aritomo12} have been applied.
In the latter model, relevant collective degrees of freedom such as the relative distance and mass partition of colliding nuclei and their deformations are chosen and non-collective degrees of freedom are treated as heat bath. 
The inertia parameter, nucleus--nucleus potential, and friction coefficient for the relevant collective degrees of freedom, appearing as transport coefficients in the equation, are important ingredients for such calculations and are mainly determined by macroscopic models.

The main motivation of the present article is to analyze the fusion hindrance
in heavy systems by \emph{microscopic} reaction models. 
In this work, we employ the time-dependent Hartree-Fock (TDHF) theory 
based on the Skyrme energy density functionals \cite{skyrme56,vautherin72,vautherin73} as a microscopic reaction model. 
Energy density functional theory is selfconsistent, fully microscopic and provides a unique tool for describing both the static and dynamical properties of nuclei across the whole nuclear chart in a unified framework \cite{bender03}.
Since the first applications of TDHF simulations to nuclear fusion reactions \cite{bonche76,koonin77,flocard78,negele78,davies78,negele82} were performed about 40 years ago,
much progress of TDHF simulations has been achieved \cite{simenel12}.
Recent three dimensional TDHF calculations are able to include full Skyrme functionals consistent with nuclear structure calculations   
and have been applied to fusion reactions \cite{kim97,simenel01, umar06, umar06a, washiyama08,washiyama09}, charge equilibration \cite{iwata10}, transfer reactions \cite{simenel10, scamps13, sekizawa13},
as well as nuclear responses \cite{simenel03, nakatsukasa05, maruhn05,avez08,ebata10,stetcu11,hashimoto12,fracasso12}.
Furthermore, TDHF simulations to reactions with heavy nuclei have been performed \cite{golabek09, kedziora10, umar10, oberacker10, guo12, simenel12, simenel14, oberacker14,wakhle14,scamps15}.
In Refs. \cite{simenel12,guo12}, extra-push energies in several heavy systems have been estimated and the fusion hindrance resulting in finite extra-push energies has been realized, which is consistent with experimental observations.
In Refs. \cite{umar10, oberacker10}, using the so-called density-constrained TDHF (DC-TDHF) method \cite{umar06a}, nucleus--nucleus potential as the minimized energy and excitation energy of a compound system are estimated along a path obtained from TDHF evolutions.
Quasifission process is extensively studied in Refs. \cite{kedziora10,oberacker14,wakhle14}, which have clarified the capability of TDHF simulations for the dynamics of quasifission.

In the present article, we analyze the fusion hindrance in detail from the point of view of energy dissipation and dynamical nucleus--nucleus potential in the entrance channel of fusion reactions in heavy systems.
For this purpose, we employ a method to simultaneously extract dynamical nucleus--nucleus potential and energy dissipation for relevant collective coordinates 
from TDHF trajectories, which we proposed in Refs. \cite{washiyama08,washiyama09} and called dissipative-dynamics TDHF (DD-TDHF) method.
The DD-TDHF method relies on the assumption that a complicated microscopic mean-field dynamics in the entrance channel of fusion reactions can be reduced to a 
macroscopic equation of motion with a dissipation term. 
Then, we apply this method to fusion reactions of heavy, nearly symmetric systems and analyze the property of nucleus--nucleus potential and energy dissipation extracted directly from TDHF simulations. 
Using the properties of the extracted potential and dissipation, we perform a systematic study of the analysis of the fusion hindrance in heavy systems.

This article is organized as follows.
In Sec.~\ref{sec:method}, we explain our method of how to analyze the fusion 
hindrance by TDHF simulations including how to extract nucleus--nucleus
potential and energy dissipation from TDHF dynamics. 
In Sec.~\ref{sec:result}, we present the results of our systematic calculations
for extra-push energies, nucleus--nucleus potentials, energy dissipation in heavy systems, and the analysis of the fusion hindrance. 
We give the conclusions of the present article in Sec.~\ref{sec:conclusions}.
We explain some details of our method in the Appendix.

\section{Method}\label{sec:method}

\subsection{Some remarks on TDHF}

In the TDHF theory, quantum many-body dynamics is replaced with 
one-body density dynamics, leading to the TDHF equation 
\begin{equation}\label{eq:TDHF}
i\hbar\frac{\partial}{\partial t}{\hat\rho}=[{\hat h}[{\hat\rho}],{\hat\rho}],
\end{equation}
where ${\hat\rho}$ and ${\hat h}[{\hat\rho}]$ denote the one-body density
and selfconsistent single-particle Hamiltonian as a functional
of one-body density, respectively. The selfconsistent single-particle 
Hamiltonian can be obtained from the energy density functional,
denoted by ${\cal E}[\hat\rho]$, through a variation ${\hat h}[{\hat\rho}]=\partial{\cal E}[\hat\rho]/\partial{\hat\rho}$.
Since the TDHF theory describes collective motions in a semiclassical way,
one can define semiclassical trajectories corresponding to the time evolutions of relevant collective coordinates from one-body density.
In this article, we concentrate on low energy nuclear dynamics, 
where the energy of the system is low enough to validate the use of the mean-field dynamics
and is above the Coulomb barrier.

\subsection{DD-TDHF method}\label{sec:method:DD-TDHF}

In this subsection, we briefly recall the DD-TDHF method, proposed in Refs. \cite{washiyama08,washiyama09}.
The method is described as follows.
First, we properly define the collective coordinates to describe fusion reactions.
In this article, we concentrate on central collisions of fusion dynamics
and employ the relative distance $R$ between projectile and target nuclei as a collective coordinate. 
An extension to using several collective coordinates is possible.
Then, we construct two-body kinematics of projectile and target nuclei.
In TDHF simulations, we define at each time 
a separation plane to separate the total density to the densities of two subsystems. The plane, called neck, is perpendicular to the collision axis, which is always $x$-axis in the case for central collisions.
We will explain how to determine the neck position in Appendix~\ref{sec:appendix:neck}.
The densities of the two subsystems are computed in the coordinate space representation as 
\begin{equation}\label{eq:subsystem}
\rho_{1,2}(\boldsymbol{r},t)\equiv \rho(\boldsymbol{r},t)\,\theta[\pm(x-x_0(t))],
\end{equation}
where $\theta(x)$ is the step function and the neck position is at $x=x_0(t)$
depending on time. 
Then, we compute the center-of-mass coordinate $R_i$ and associated momentum $P_i$ as well as the mass number $A_i$ for each subsystem $i=1,2$ from the density of each subsystem, $A_i(t)={\rm Tr}(\hat{\rho}_i(t))$, $R_i(t)={\rm Tr}(\hat{x} \hat{\rho}_i(t)) / A_i$, and $P_i(t)={\rm Tr}(\hat{p}_x \hat{\rho}_i(t))$. 
The masses of the subsystems are directly computed as 
\begin{equation}
m_{i} = P_{i}/\dot{R}_{i}, 
\end{equation}
for $i=1,2$.
From these quantities, we construct a TDHF trajectory for the relative distance $R$, associated momentum $P$, and reduced mass $\mu$ at each time by
\begin{align}
R& = R_1 - R_2, \\
P& = \frac{m_2 P_1 - m_1 P_2}{m_1 + m_2}, \\
\mu& = \frac{m_1 m_2}{m_1 + m_2}. \label{eq:reduced-mass}
\end{align}
We assume that the time evolutions of $R$ and $P$ obtained from the TDHF evolution of central collisions follow a one-dimensional macroscopic equation of motion with a dissipation term that is assumed to depend on the velocity ${\dot R}$:
\begin{align}
\frac{dR}{dt}& =  \frac{P}{\mu}, \label{eq:newton1} \\
\frac{dP}{dt}& = -\frac{dV}{dR} -\frac{d}{dR}\left(\frac{P^2}{2\mu}\right) 
                 -\gamma \frac{dR}{dt}, \label{eq:newton2}
\end{align}
where $V$ and $\gamma$ denote the nucleus--nucleus potential and 
friction coefficient, respectively.
This assumption is based on the hypothesis that 
a complicated microscopic dynamics obtained from the TDHF evolution 
can be reduced to a simple one-dimensional Newton equation.
The aim is to obtain transport coefficients in the equation from TDHF dynamics. 
We have already obtained the reduced mass by Eq.~(\ref{eq:reduced-mass}).
To obtain two unknown quantities $dV/dR$ and $\gamma$ as a function of $R$, 
we need two equations.
If we assume that the transport coefficients, $\mu$, $dV/dR$, and $\gamma$, at each $R$ do not change by a slight change of energy $\Delta E$,
we can construct a system of two equations for Eq.~(\ref{eq:newton2}) 
corresponding to two TDHF trajectories at two slightly different energies $E_I$ and $E_{II}=E_I+\Delta E$. 
By solving the system of two equations at $R$, we obtain $\gamma(R)$ and $dV/dR$ as
\begin{align}
\gamma(R)&=-\frac{\dot{P}_I-\dot{P}_{II}}{\dot{R}_I-\dot{R}_{II}}+\frac{\dot{R}_{I}+\dot{R}_{II}}{2}\frac{d\mu}{dR}, \label{eq:gamma} \\
\frac{dV(R)}{dR}&=\frac{\dot{R}_I\dot{P}_{II}-\dot{R}_{II}\dot{P}_I}{\dot{R}_I-\dot{R}_{II}}-\frac{\dot{R}_{I}\dot{R}_{II}}{2}\frac{d\mu}{dR},\label{eq:dvdr}
\end{align}
where the subscripts $I$ and $II$ correspond to the trajectories at energies $E_{I}$ and $E_{II}$, respectively, 
and we perform all the time derivatives at $R_{I}=R_{II}=R$.
The nucleus--nucleus potential $V$ is obtained from Eq.~(\ref{eq:dvdr}) by integration with the asymptotic form of the Coulomb potential at sufficiently large distances. We employ $R_{\rm max}=20$\,fm in this article.
Therefore, we do not need any normalizations for the potential.
In Refs. \cite{washiyama08,washiyama09}, we applied the DD-TDHF method to fusion reactions in light- and medium-mass systems and gave the following results: 
(i) At energies near the Coulomb barrier, dynamical effects significantly decrease the barrier height of the extracted nucleus--nucleus potentials because of reorganization of the density of colliding nuclei, resulting in energy dependence of potential.
(ii) The extracted friction coefficients show a universal property similar to that in one-body dissipation models, and significant energy dependence
related to the picture of the window formula of one-body dissipation mechanism.
From those results, we concluded that the DD-TDHF method is useful tool to investigate the property of nucleus--nucleus potential and one-body energy dissipation in low-energy nuclear reactions.

When we apply the DD-TDHF method to fusion reactions in heavy systems,
we find that, as the overlap of the densities of colliding nuclei becomes substantially large at small relative distances in TDHF simulations, 
constructing two-body kinematics of colliding nuclei with a proper determination of the neck position becomes difficult.
Also, we face in some systems the following situations:
(i) the relative velocity or the velocity of a fragment becomes 0, 
(ii) the mass of a fragment becomes negative.
We consider that these situations also make it impossible to construct two-body kinematics.
Therefore, we stop performing the extraction once we face these situations or once we can not define the neck position.
In the following, we denote as $R_{\rm min}$ the relative distance at which we stop the extraction.

\subsection{Nucleus--nucleus potential from the frozen density approximation}
\label{sec:method:FD}

As a reference for the nucleus--nucleus potential based on a framework of 
the energy density functional, we employ the so-called frozen density approximation \cite{brueckner68,denisov02}.
In this approximation, the energy of the total system at a given relative
distance $R$ between the centers-of-mass of the projectile and target densities, denoted by ${\cal E}[\rho_{P+T}](R)$, is obtained 
with keeping their densities frozen to their respective ground-state densities. 
Denoting the ground-state densities of the projectile and target nuclei as $\rho_P$ and $\rho_T$, respectively, we obtain the frozen density potential as 
\begin{equation}\label{eq:FD}
V^{\rm FD}(R) = {\cal E}[\rho_{P+T}](R) - {\cal E}[\rho_P] - {\cal E}[\rho_T],
\end{equation}
where ${\cal E[\rho]}$ is calculated from the same Skyrme energy density functionals as those used in our TDHF simulations.
In the frozen density potential, effects of possible rearrangement of internal structure during collision are not considered.
Note that we neglect the Pauli effect from the overlap of the projectile and target densities in 
estimating ${\cal E}[\rho_{P+T}](R)$. Namely, we approximate $\rho_{P+T}$ as a simple sum of $\rho_P+\rho_T$.
This will be crucial when the overlap becomes large at small relative distances,
leading to overestimation of the attractive potential energy.
Therefore, we use the frozen density potential as a reference 
until a relative distance around the Coulomb barrier radius,
where the overlap of projectile and target densities is expected to be small.

\subsection{Definition of fusion in heavy systems}\label{sec:method:definition}

Semiclassical nature of TDHF dynamics enables us to define a threshold energy for classification of nuclear reactions. 
In the case of central collisions at low energies, 
we can define the fusion threshold energy above which fusion, i.e., the formation of a compound nucleus, takes place.
Following a manner similar to those in previous works \cite{simenel12,guo12},
we define fusion as a reaction where a colliding system keeps
a compact shape for a sufficiently long time ($\sim 2000$\,fm/$c$)
after contact. 
From a sufficiently low energy, we perform TDHF calculations systematically by
increasing progressively the energy with a step of 0.5\,MeV until fusion is reached.

We have found that there is an ambiguity to define a compact shape
leading to the formation of the compound nucleus in some systems.
We will come to this point in Appendix~\ref{sec:appendix:threshold},
and will show that this ambiguity does not affect the conclusion of this article.

\subsection{Numerical setups}\label{sec:method:setup}

In this work, we perform only central collisions.
In our practical computations of TDHF evolutions,
we use the TDHF3D code \cite{kim97} developed by P.~Bonche and coworkers
with the SLy4d parametrization \cite{kim97} of the Skyrme energy density functional, 
which is slightly modified from the SLy4 parametrization \cite{chabanat97,chabanat98} by neglecting the center-of-mass corrections in the fitting protocol of determining the parameter set.
As the initial conditions, we solve static Hartree-Fock equations by using the computer code ev8 \cite{ev8} with the imaginary time method. 
Then, we place the centers of the projectile and target densities on the $x$-axis.
The initial distance is set to be $R_0=28.8$\,fm for all the systems
considered in this article except for the $^{40}$Ca\,+\,$^{40}$Ca system ($R_0=22.4$\,fm).
We assume that the colliding nuclei follow the Rutherford trajectory 
before the initial distance.
The step sizes in the coordinate space and time are 0.8\,fm and 0.45\,fm/$c$, respectively, for both static and dynamical computations.
The numbers of the mesh points in our three dimensional numerical box is $64\times28\times14$ along the $x$, $y$, and $z$-axis, respectively, with a symmetry plane at $z=0$. 
The pairing correlation is neglected in both static and dynamical calculations, though this has been included in 
some of recent applications to nuclear responses and reactions in 
either time-dependent Hartree-Fock-Bogoliubov \cite{avez08,stetcu11,hashimoto12} 
or an approximated way \cite{ebata10,scamps13}.

\section{Results}\label{sec:result}

To analyze the fusion hindrance, we perform TDHF simulations for central collisions in heavy systems
$^{90}$Zr\,+\,$^{90,92,94,96}$Zr,  $^{100}$Mo\,+\,$^{92,100}$Mo,$^{104}$Ru,$^{110}$Pd,
and $^{124}$Sn\,+\,$^{90,92,94,96}$Zr, where the extra-push energies were deduced
from experimental evaporation-residue cross sections \cite{sahm85, schmidt91},
and $^{96}$Zr\,+\,$^{132}$Sn,$^{136}$Xe, $^{70}$Zn\,+\,$^{208}$Pb.
 
We obtain the ground-state shapes of $^{92}$Zr, $^{92}$Mo, $^{100}$Mo, and $^{124}$Sn to be oblate with the deformation parameters $\beta=0.03$, $0.05$, $0.23$, and $0.11$, respectively.
The ground-state shapes of $^{94}$Zr and $^{136}$Xe are slightly prolate deformation with $\beta=0.04$ and $0.05$, respectively.
The ground-states of $^{104}$Ru and $^{110}$Pd are triaxial with $\beta=0.31$, $\gamma=21.0^\circ$ and $\beta=0.31$, $\gamma=21.6^\circ$, respectively.
The other nuclei are found to be spherical.
Note that the ground-states of $^{92,94}$Zr, $^{92}$Mo, $^{124}$Sn, and $^{136}$Xe
would be spherical if the pairing correlation were included in the calculations.
For collisions with deformed nuclei, we place the nucleus in the box by setting the symmetry axis of the axially deformed nucleus or the longest axis of the triaxial nucleus parallel to the $z$-axis.

\begin{figure}[tb]
\includegraphics[width=0.9\linewidth,clip]{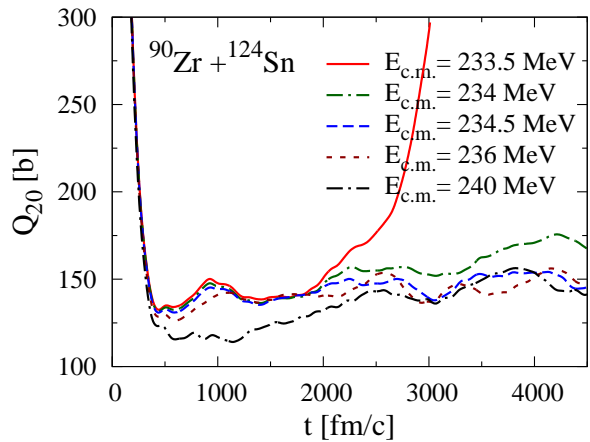}
\caption{\label{fig:q20Zr90Sn124} (Color online)
Time evolution of mass quadrupole moment $Q_{20}$ of the total system 
at different center-of-mass energies for the $^{90}$Zr\,+\,$^{124}$Sn system.
}
\end{figure}
\begin{figure}[tb]
\includegraphics[width=\linewidth,clip]{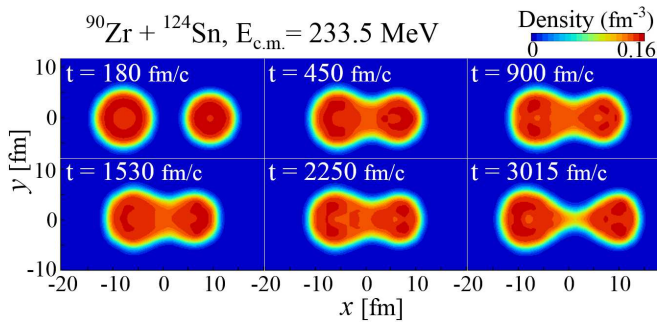}
\caption{\label{fig:2DdensityZr90Sn124} (Color online)
Time evolution of density profiles $\rho(x,y,z=0,t)$ at $E_{\rm c.m.}=233.5$\,MeV 
for the $^{90}$Zr\,+\,$^{124}$Sn system.
}
\end{figure}

\begin{figure*}[tb]
\includegraphics[width=0.9\linewidth,clip]{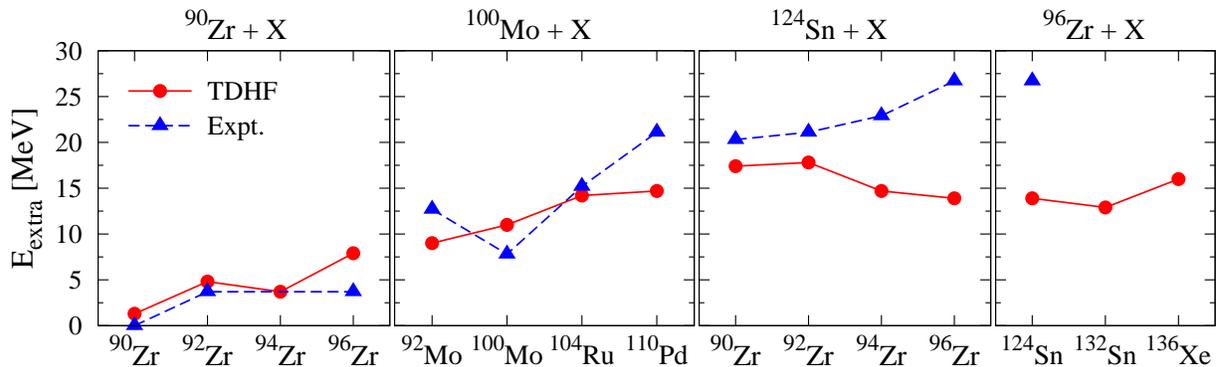}
\caption{\label{fig:extrapush} (Color online)
Extra-push energies extracted from TDHF simulations (red filled circles) and from
experimental data (blue filled triangles) \cite{schmidt91}.
In the $x$-axis, the nuclide of the target is shown, while the nuclide of the projectile is shown at the top.
}
\end{figure*}

\begin{figure}[tb]
\includegraphics[width=0.9\linewidth,clip]{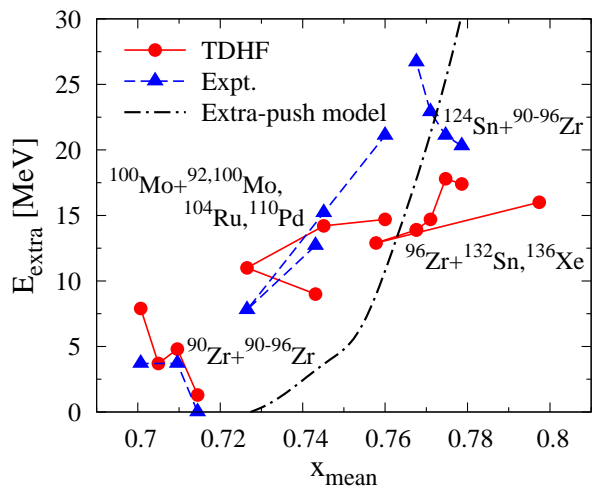}
\caption{\label{fig:extrapush.fissility} (Color online)
Same as in Fig.~\ref{fig:extrapush}, but as a function of parameter $x_{\rm mean}$ defined by Eq.~(\ref{eq:mean_fissility}).}
\end{figure}

\subsection{Extra-push energy}\label{sec:result:extrapush}

In our TDHF simulations in heavy systems, 
we indeed observe that, after contact of colliding nuclei,  
a colliding system keeps a configuration of substantial overlap of projectile and target densities for a long time, and then reseparates to two fragments.
We regard this process as quasifission. 
This has been observed in previous works \cite{simenel12,kedziora10,guo12} 
and we are not aware of such a process in lighter systems.
As an example of quasifission realized in our TDHF simulations,
we show in Fig.~\ref{fig:q20Zr90Sn124} the time evolution of mass quadrupole moment of the $^{90}$Zr\,+\,$^{124}$Sn system defined as 
\begin{equation}\label{eq:q20}
Q_{20}(t) = \int d^3r \, (2x^2-y^2-z^2)\, \rho(x,y,z,t),
\end{equation}
where $\rho(x,y,z,t)$ denotes the density of the total system obtained from TDHF calculations at different center-of-mass energies $E_{\rm c.m.}$ 
In this system, we can identify that the case at $E_{\rm c.m.}=233.5$\,MeV leads to quasifission after a long contact time $\sim 1800$\,fm/$c$.
This energy is well above the barrier height obtained from 
the frozen density potential, 217.1\,MeV.
In Fig.~\ref{fig:2DdensityZr90Sn124}, we show the time evolution of density profile for the quasifission process at $E_{\rm c.m.}=233.5$\,MeV. 
Large overlap of the densities from $t=450$\,fm/$c$ to $t=2250$\,fm/$c$ and reseparation at $t=3015$\,fm/$c$ can be seen.

We define the extra-push energy $E_{\rm extra}$ estimated from TDHF simulations as
\begin{equation}
E_{\rm extra}= E_{\rm thres} -V_{\rm FD},
\end{equation}
where $E_{\rm thres}$ and $V_{\rm FD}$ denote the fusion threshold energy and 
the potential barrier height obtained from the frozen density approximation, respectively.
Figure~\ref{fig:extrapush} shows the extra-push energies obtained from TDHF simulations denoted by the red filled circles for various heavy systems.
We also show by the blue filled triangles experimental extra-push energies taken from Ref. \cite{schmidt91}, 
which are defined as the difference between the apparent barrier deduced from experimental data of evaporation-residue cross sections and the barrier height obtained from the Bass model \cite{bass74}.
Our TDHF simulations reasonably reproduce the experimental extra-push energies,
except for heavier systems.
We also reproduce the trend that the extra-push energy becomes large as the charge product of the system becomes large.
Figure~\ref{fig:extrapush.fissility} compares our result with the extra-push model \cite{swiatecki81,swiatecki82,bjornholm82,blocki86}, plotted by the dot-dashed line. 
In this figure, we plot extra-push energy as a function of parameter $x_{\rm mean}$ defined as
\begin{equation}\label{eq:mean_fissility}
x_{\rm mean}= \frac{1}{3}x_{\rm eff}+\frac{2}{3}x_{\rm fis}
\end{equation}
with 
\begin{align}
x_{\rm eff}&=\frac{4Z_PZ_T}{A_P^{1/3}A_T^{1/3}\left(A_P^{1/3}+A_T^{1/3}\right)}
\biggr/\left(\frac{Z^2}{A}\right)_{\rm crit}, \label{eq:fissility} \\
x_{\rm fis}&=\left(Z^2/A\right)\Bigr/\left(Z^2/A\right)_{\rm crit}. 
\label{eq:fissility1}
\end{align}
Here, $(Z^2/A)_{\rm crit}= 50.883(1 - 1.7826I^2)$ is used, where $I=(A-2Z)/A$ is the neutron excess of the compound nucleus with $A=A_P+A_T$, $Z=Z_P+Z_T$ being the mass and charge numbers of the compound nucleus, respectively.
The parameter $x_{\rm mean}$ was originally proposed in Ref. \cite{blocki86}, 
with which the extra-push energy better scales.
Around $x_{\rm mean}=0.74$, our result is better than the extra-push model,
while the extra-push model is better at $x_{\rm mean}\sim 0.77$.

In some systems, we find that our TDHF simulations underestimate the values of the experimental extra-push energy. 
We underestimate the extra-push energy by about 12\,MeV in the $^{96}$Zr\,+\,$^{124}$Sn system. 
A possible interpretation of this underestimation is as follows.
As we have explained in Sec.~\ref{sec:method:definition},
we define fusion by TDHF as a reaction where a colliding system keeps
a compact shape for a sufficiently long time. 
As one can see in Fig.~\ref{fig:q20Zr90Sn124},  
such a reaction never reaches to the shape of resulting compound nucleus in TDHF simulations
because of lack of dissipation beyond the mean-field level, 
e.g., two-body dissipation arising from direct nucleon--nucleon collisions \cite{wen13,jiang13,wen14}.
Therefore, the cases defined as fusion from TDHF simulations might lead to
reseparation (quasifission) if dissipation mechanisms beyond the mean-field level were included in the calculations.
That is, fusion threshold energies in some systems might be underestimated within the mean-field level.
Other possible interpretation of this underestimation is the neglect of pairing correlations in both static and dynamical calculations. This might induce the discrepancy between TDHF and experimental fusion threshold energies.

Although we have such a limitation, 
our TDHF simulations qualitatively reproduce the overall property of 
the dependence of observed extra-push energies on system size.
This fact validates our following analysis based on TDHF simulations.
In the following, we will give detailed analyses of nucleus--nucleus
potential and energy dissipation obtained from the DD-TDHF method 
and of the origin of the fusion hindrance in heavy systems.

\subsection{Nucleus--nucleus potential in heavy systems}
\label{sec:result:potential}

\subsubsection{Property of extracted potential}

\begin{figure}[tb]
\includegraphics[width=0.9\linewidth,clip]{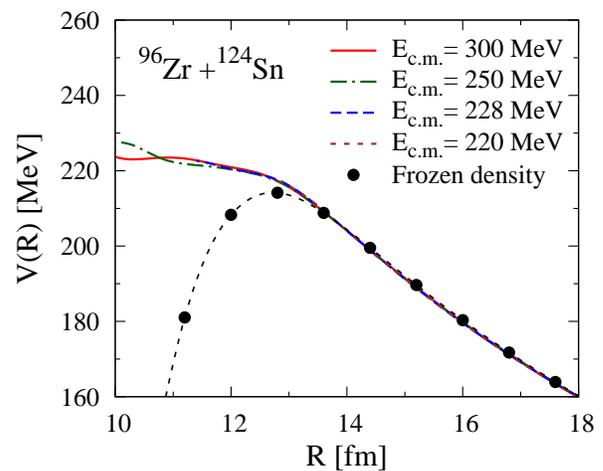}
\caption{\label{fig:potZr96Sn124} (Color online)
Nucleus--nucleus potential as a function of relative distance $R$ for the $^{96}$Zr\,+\,$^{124}$Sn system. 
The red solid, green dot-dashed, blue dashed, and brown dotted lines denote the potentials extracted from the TDHF trajectories at $E_{\rm c.m.}=300$, 250, 228, and 220\,MeV, respectively.
The filled circles are the potential obtained from the frozen density approximation.}
\end{figure}
\begin{figure}[tb]
\includegraphics[width=0.8\linewidth,clip]{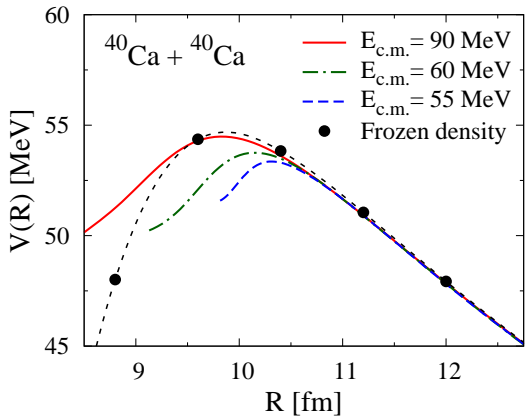}
\caption{\label{fig:potCa40Ca40} (Color online)
Same as in Fig.~\ref{fig:potZr96Sn124}, but for the $^{40}$Ca\,+\,$^{40}$Ca system.
}
\end{figure}

Figure~\ref{fig:potZr96Sn124} shows the nucleus--nucleus potentials
as a function of relative distance $R$
obtained from the DD-TDHF method for the $^{96}$Zr\,+\,$^{124}$Sn system at $E_{\rm c.m.}=300$, 250, 228, and 220\,MeV used in TDHF. The filled circles are the potential obtained by the frozen density approximation. 
Note that, when $R \le 11.2$\,fm, the overlap density of the two nuclei at the neck in the frozen density approximation becomes close to the normal density 0.16\,fm$^{-3}$, indicating the violation of the frozen density approximation at $R \le 11.2$\,fm.
The charge product of this system is 2000, and we observe the fusion hindrance in this system as we have shown in Fig.~\ref{fig:extrapush}.
The fusion threshold energy for this system is $E_{\rm c.m.}=228$\,MeV.
At $E_{\rm c.m.}=220$\,MeV, which is higher than the barrier height of the frozen density potential, the system reseparates after slight contact.
Note that, for the cases at $E_{\rm c.m.}=250$, 228, and 220\,MeV, we stop the extraction of the potential at $R_{\rm min}=10.1$, 11.4, and 12.7\,fm, respectively.
The extracted potentials agree with the frozen density potential until $R\sim 13.5$\,fm and start to deviate from the frozen density one as $R$ decreases.
At $R< 13.5$\,fm, the extracted potentials do not have an
ordinary barrier structure, though the frozen density potential has an ordinary barrier at $R\sim 12.8$\,fm.
The extracted potentials monotonically increase after deviating from 
the frozen density potential.
At $R > 11.4$\,fm, the extracted potentials agree with each other, while a deviation is seen between the potentials extracted at $E_{\rm c.m.}=300$ and 250\,MeV
at $R < 11.4$\,fm.
To investigate these findings in detail, 
we compare them with the potentials extracted in light systems.
Figure~\ref{fig:potCa40Ca40} show the potentials extracted in the $^{40}$Ca\,+\,$^{40}$Ca system.
From the comparison, we find two main differences between the potentials extracted in light and heavy systems as follows:
\begin{itemize}
\item Potentials for the $^{40}$Ca\,+\,$^{40}$Ca system and the frozen
density potential for both systems have an ordinary barrier structure, 
while no barrier appears in the potentials extracted in the $^{96}$Zr\,+\,$^{124}$Sn system.
Monotonic increase of the extracted potentials is only observed in the $^{96}$Zr\,+\,$^{124}$Sn system.
\item In the $^{40}$Ca\,+\,$^{40}$Ca system, energy dependence of potential around the Coulomb barrier is significant, 
while for the $^{96}$Zr\,+\,$^{124}$Sn system, energy dependence of potential is less pronounced than in the $^{40}$Ca\,+\,$^{40}$Ca system.
\end{itemize}

\begin{figure}[tb]
\includegraphics[width=\linewidth,clip]{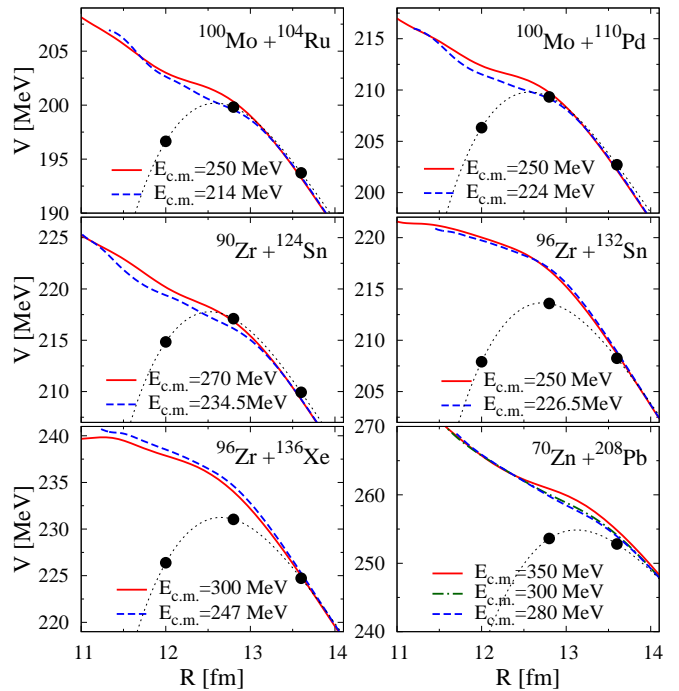}
\caption{\label{fig:potall} (Color online)
Nucleus--nucleus potentials for various heavy systems are displayed 
at the region around the barrier of the frozen density potential.
}
\end{figure}

The above two properties are indeed general in heavy systems.
To see this, we show in Fig.~\ref{fig:potall} extracted nucleus--nucleus potentials for the $^{100}$Mo\,+\,$^{104}$Ru ($Z_PZ_T=1848$), $^{100}$Mo\,+\,$^{110}$Pd (1932), $^{90}$Zr\,+\,$^{124}$Sn, $^{96}$Zr\,+\,$^{132}$Sn (2000), $^{96}$Zr\,+\,$^{136}$Xe (2160), $^{70}$Zn\,+\,$^{208}$Pb (2460) systems.
In each system, except for the $^{70}$Zn\,+\,$^{208}$Pb system, we show the potentials extracted at the fusion threshold energy 
by the blue dashed line and at higher energy by the red solid line.
In the $^{70}$Zn\,+\,$^{208}$Pb system, we do not observe fusion at energies
between 265 and 400\,MeV.
In any systems, the potentials show monotonic increase as $R$ decreases
and energy dependence of potential is less pronounced.
(At most, the difference is 0.6\% in the $^{90}$Zr\,+\,$^{124}$Sn system, while the difference is 3\% in the $^{40}$Ca\,+\,$^{40}$Ca system.)
By this systematic study of nucleus--nucleus potential, 
we conclude that these properties of the extracted potentials are general in heavy systems.

\subsubsection{Discussion}

\begin{figure}[tb]
\includegraphics[width=0.8\linewidth,clip]{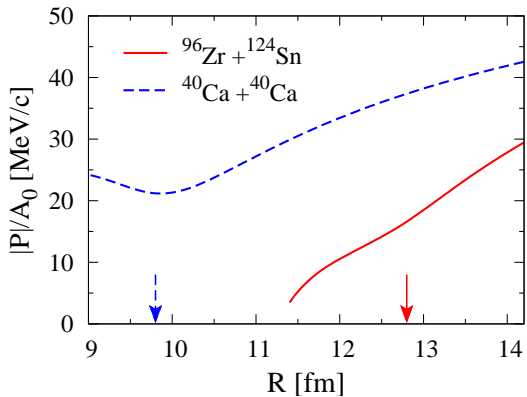}
\caption{\label{fig:momentum} (Color online)
Absolute value of the relative momentum scaled by the initial reduced mass number $A_0$ as a function of relative distance for the $^{96}$Zr\,+\,$^{124}$Sn at $E_{\rm c.m.}=228$\,MeV (red solid line) and the $^{40}$Ca\,+\,$^{40}$Ca at $E_{\rm c.m.}=60$\,MeV (blue dashed line) systems.
The barrier radii of the frozen density potentials for these systems are indicated by arrow.
}
\end{figure}

We have shown that the nucleus--nucleus potentials extracted in heavy systems exhibit monotonic increase as the relative distance decreases and do not have an ordinary barrier structure and this property is seen only in heavy systems. 
In the following, we will consider the reason why this property appears only in this case.

Figure~\ref{fig:momentum} shows the relative momentum scaled by the initial reduced mass number $A_0=A_PA_T/(A_P+A_T)$ as a function of relative distance 
for the $^{96}$Zr\,+\,$^{124}$Sn (red solid line) and the $^{40}$Ca\,+\,$^{40}$Ca (blue dashed line) systems. 
The Coulomb barrier radius of the frozen density potential is indicated by arrow for these systems.
Note that the sign of the relative momentum is negative in the entrance channel of central collisions.
In the $^{40}$Ca\,+\,$^{40}$Ca system, the momentum decreases as $R$ decreases until the Coulomb barrier, and then the momentum starts to increase after the system overcomes the Coulomb barrier.
In the $^{96}$Zr\,+\,$^{124}$Sn system, however, the momentum monotonically decreases as $R$ decreases.
If the potential of this system were the frozen density one, the momentum would increase after passing the barrier. Therefore, the monotonic decrease of the momentum with $R$ decreasing induces the increase in extracted potentials at the inside of the frozen density barrier. 
Dissipation also can decrease the momentum.
However, our analysis shows that the extracted dissipation is not enough to explain the property of the time evolution of the momentum in heavy systems and the increase in potential is necessary. 

\begin{figure}[tb]
\includegraphics[width=0.8\linewidth,clip]{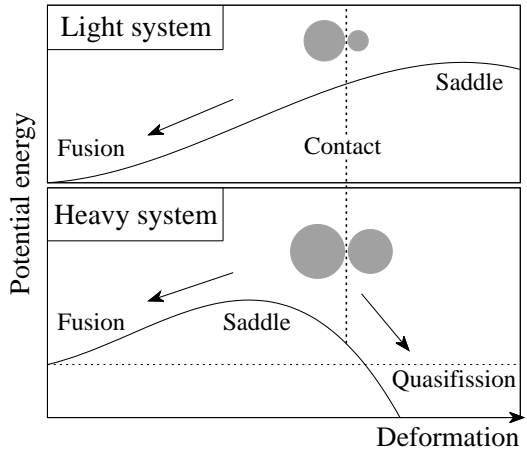}
\caption{\label{fig:conditional_saddle} 
Schematic illustration of the appearance of a conditional saddle in heavy systems. See text for details.
}
\end{figure}

We further consider 
the meaning of  
the increase in potential at small relative distances. To do so, we introduce with Fig.~\ref{fig:conditional_saddle}
an explanation of the appearance of the fusion hindrance in heavy systems
from a simple geometrical feature of potential used in the extra-push model \cite{swiatecki81,swiatecki82,bjornholm82}. 
Figure~\ref{fig:conditional_saddle} shows the schematic illustration of an adiabatic potential energy landscape as a function of deformation of a compound nucleus in light and heavy systems.
In both systems, the left region corresponds to the configuration of a compound nucleus formation, while the right corresponds to the quasifission. 
The potential barrier, denoted by ``saddle'' in the figure, is usually called conditional saddle and corresponds to the fission barrier under the condition that the mass partition of the system into two fragments is fixed to be the one at the contact configuration in the entrance channel.
If the equilibration at the contact configuration is fast enough, in other words, the transition from two-body regime to one-body regime in the collision dynamics is fast enough, the system will feel this adiabatic potential once two nuclei touch.
In light systems, the contact configuration is located inside the saddle on the potential landscape. Consequently, fusion automatically takes place once the two nuclei touch.
The situation is changed for heavy systems because of large Coulomb energy. 
That is, the saddle is shifted to the inside of the contact configuration,
and the system needs to overcome the saddle for fusion, which gives rise to the fusion hindrance in the extra-push model.
In the DD-TDHF method, dynamical effects are automatically included in the extracted potential through TDHF dynamics. 
Since the TDHF theory treats the reorganization of single-particle wave functions during collision in a selfconsistent way, 
the extracted potential includes effects of smooth transition from two-body regime of collision dynamics to one-body regime of shape evolution at small $R$.
Therefore, we consider that the increase in potential at small $R$ observed in heavy systems can be regarded as a reflection of the appearance of the conditional saddle inside the contact configuration in the schematic picture explained above.

\subsubsection{Isotope dependence of nucleus--nucleus potential}

\begin{figure}[tb]
\includegraphics[width=0.9\linewidth,clip]{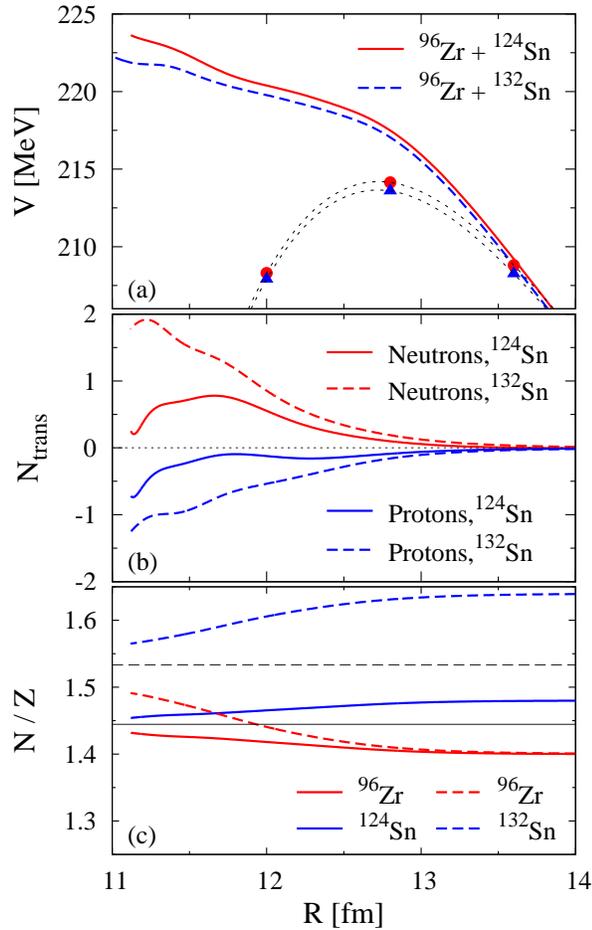}
\caption{(Color online)
\label{fig:potZr96Sn}(a)  
Comparison between the nucleus--nucleus potentials for the $^{96}$Zr\,+\,$^{124}$Sn (red solid line) and $^{96}$Zr\,+\,$^{132}$Sn (blue dashed line) systems. Both potentials are extracted at $E_{\rm c.m.}=230$\,MeV.
The red filled circles and blue filled triangles are the frozen density potentials for the $^{96}$Zr\,+\,$^{124}$Sn and $^{96}$Zr\,+\,$^{132}$Sn systems, respectively.
\label{fig:transZr96Sn}(b) 
Average number $N_{\rm trans}$ of transferred neutrons (red) and protons (blue) from $^{124}\mathrm{Sn}$ (solid line) and $^{132}\textrm{Sn}$ (dashed line) to $^{96}$Zr in the $^{96}$Zr\,+\,$^{124,132}$Sn systems at $E_{\rm c.m.}=230$\,MeV.
(c)\label{fig:NZratioZr96Sn}
Neutron to proton ratio ($N/Z$) of the two subsystems in the $^{96}$Zr\,+\,$^{124}$Sn (solid line) systems and $^{96}$Zr\,+\,$^{132}$Sn (dashed line) systems.
The gray solid and dashed lines indicate the values of the equilibrated $N/Z$ for the two systems.
}
\end{figure}

Next we will analyze the isotope dependence 
of nucleus--nucleus potentials in heavy systems, 
For this purpose, we employ the $^{96}$Zr\,+\,$^{124}$Sn and $^{96}$Zr\,+\,$^{132}$Sn systems by changing the neutron number of Sn.

In Fig.~\ref{fig:potZr96Sn}(a), we show the nucleus--nucleus potentials 
for the $^{96}$Zr\,+\,$^{124}$Sn (red solid line) and $^{96}$Zr\,+\,$^{132}$Sn (blue dashed line) systems. 
The potentials in both systems are extracted at $E_{\rm c.m.}=230$\,MeV.
As a reference, the frozen density potentials are plotted by the red filled circles and blue filled triangles for the $^{96}$Zr\,+\,$^{124}$Sn and $^{96}$Zr\,+\,$^{132}$Sn systems, respectively.
In the DD-TDHF method and frozen density approximation, the potential energy for the $^{96}$Zr\,+\,$^{124}$Sn system is slightly larger than that for the $^{96}$Zr\,+\,$^{132}$Sn system.
In the DD-TDHF case, the difference between the potentials is 0.4\,MeV at the barrier radius of the frozen density potential, $R=12.8$\,fm, which is similar to the difference between the frozen density potential barriers of the two systems, 0.5\,MeV.
Then, the difference between the two extracted potentials evolves as the relative distance becomes small, and becomes 1.8\,MeV at $R=11.1$\,fm.
In order to further understand the role of the neutron excess in $^{132}$Sn on the potential, we show in Fig.~\ref{fig:transZr96Sn}(b) the average numbers of transferred nucleons, $N_{\rm trans}$, through the neck from $^{124}$Sn and $^{132}$Sn to $^{96}$Zr in the $^{96}$Zr\,+\,$^{124,132}$Sn systems at $E_{\rm c.m.}=230$\,MeV. 
The positive and negative values correspond to transfer from $^{124,132}$Sn to $^{96}$Zr and transfer from $^{96}$Zr to $^{124,132}$Sn, respectively.
We find that the number of transferred neutrons in the $^{96}$Zr\,+\,$^{132}$Sn system is larger than that in the $^{96}$Zr\,+\,$^{124}$Sn system as $R$ decreases. 
We also find the larger number of transferred protons from $^{96}$Zr to $^{132}$Sn 
than that from $^{96}$Zr to $^{124}$Sn. 
Because of larger neutron to proton ($N/Z$) ratio of $^{132}$Sn ($N/Z=1.64$) than that of $^{124}$Sn ($N/Z=1.48$), 
larger nucleon transfer towards charge equilibration occurs during the collision in the $^{96}$Zr\,+\,$^{132}$Sn system.
This is illustrated in Fig.~\ref{fig:NZratioZr96Sn}(c), where $N/Z$ of the two subsystems in the $^{96}$Zr\,+\,$^{124}$Sn (solid line) and $^{96}$Zr\,+\,$^{132}$Sn (dashed line) systems are plotted. 
As $R$ decreases, larger change in $N/Z$ towards the change equilibration
in the $^{96}$Zr\,+\,$^{132}$Sn system can be seen.  
We consider that the net effects make the potential for the $^{96}$Zr\,+\,$^{132}$Sn system lower.

\begin{figure}[tb]
\includegraphics[width=0.9\linewidth,clip]{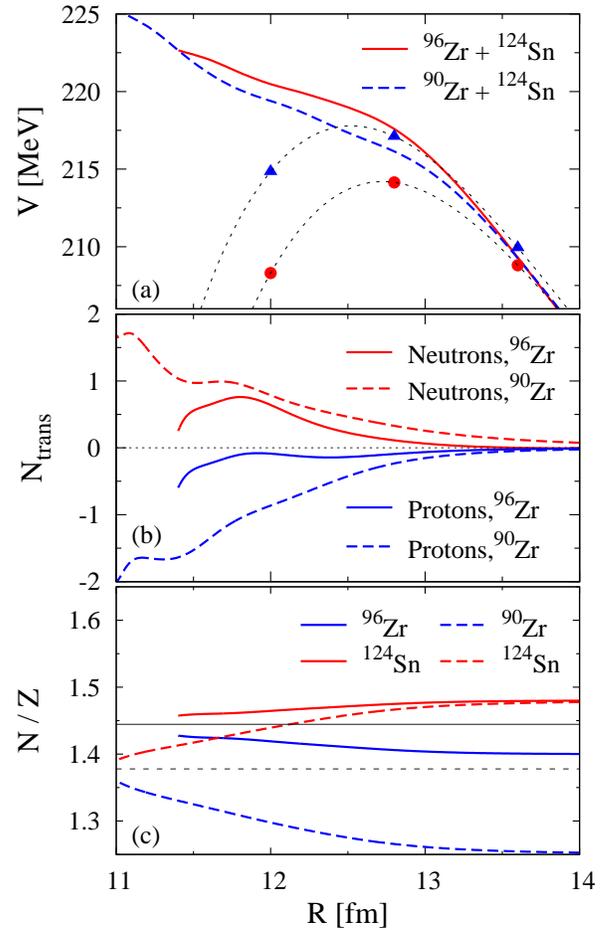}
\caption{\label{fig:ZrSn124}(Color online)
Same as Fig.~\ref{fig:potZr96Sn} but for the $^{90,96}$Zr\,+\,$^{124}$Sn systems
obtained at the fusion threshold energy $E_{\rm thres}=234.5$\,MeV and $E_{\rm thres}=228$\,MeV, respectively.
}
\end{figure}

Similar property is found in the extracted potentials in $^{90,96}$Zr\,+\,$^{124}$Sn systems.
Figure~\ref{fig:ZrSn124}(a) shows the extracted potentials at the fusion threshold energy and the frozen density potentials for these systems. We find that the increase of the extracted potential from the frozen density potential is larger in $^{96}$Zr\,+\,$^{124}$Sn than in $^{90}$Zr\,+\,$^{124}$Sn. The difference is about 4\,MeV in average.
From Fig.~\ref{fig:ZrSn124}(b) showing the number of transferred nucleons
and Fig.~\ref{fig:ZrSn124}(c) showing $N/Z$ ratio,
larger nucleon transfer towards charge equilibration occurs in the $^{90}$Zr\,+\,$^{124}$Sn system due to larger difference between the initial $N/Z$ ratios of the projectile and target.

From these observations, 
we conclude that, as nucleon transfer towards charge equilibration during the collision becomes smaller, increase in the extracted potential from the frozen density potential becomes larger.

\subsubsection{Comparison with the DC-TDHF method}\label{sec:result:DCTDHF}

\begin{figure}[tb]
\includegraphics[width=0.9\linewidth,clip]{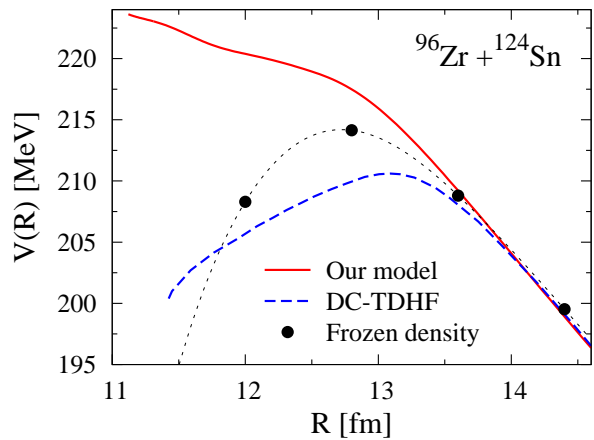}
\caption{\label{fig:potZr96Sn124compare} (Color online) 
Comparison between the nucleus--nucleus potential obtained from our method (red solid line) and that from the DC-TDHF method (blue dashed line) taken from Ref. \cite{oberacker10} for the $^{96}$Zr\,+\,$^{124}$Sn system at $E_{\rm c.m.}=230$\,MeV.
As a reference, the frozen density potential of this system is plotted by the filled circles.
}
\end{figure}

Figure~\ref{fig:potZr96Sn124compare} compares nucleus--nucleus potentials obtained from our method (red solid line) and from the DC-TDHF method (blue dashed line) in Ref. \cite{oberacker10} for the $^{96}$Zr\,+\,$^{124}$Sn system at $E_{\rm c.m.}=230$\,MeV,
which is 2-MeV above the fusion threshold energy.
As a reference, the frozen density potential is plotted by the filled circles.
The DC-TDHF potential has an ordinary barrier at $R\sim 13.1$\,fm,
which is slightly outside the barrier radius of the frozen density potential, $R\sim 12.8$\,fm. 
The barrier height of the DC-TDHF potential is about 3.5-MeV lower than that of the frozen density potential.
The difference between DC-TDHF potential and our potential becomes
large as the relative distance becomes small.

\begin{figure}[tb]
\includegraphics[width=0.8\linewidth,clip]{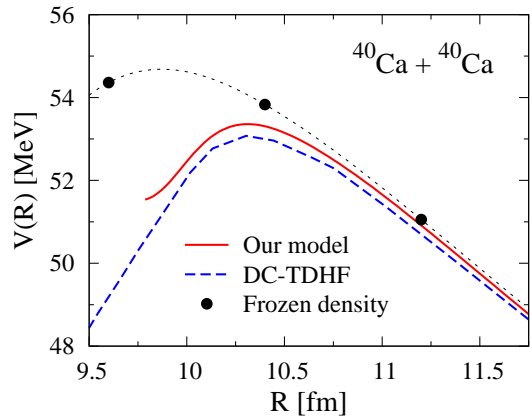}
\caption{\label{fig:potCa40Ca40compare} (Color online) 
Same as in Fig.~\ref{fig:potZr96Sn124compare} but for the $^{40}$Ca\,+\,$^{40}$Ca system at $E_{\rm c.m.}=55$\,MeV.
DC-TDHF potential is taken from Ref. \cite{umar14}. 
}
\end{figure}

Since both methods solve the same TDHF equations,
the time evolutions of the single-particle wave functions in the two methods should be same, except for numerical errors and slightly different Skyrme's parametrizations (SLy4d for DD-TDHF and SLy4 for DC-TDHF).
In the DC-TDHF method, the potential energy is given as the energy density functional of the total system with the minimization under the constraint that the total density matches the TDHF density.
Therefore, the obtained potential can be regarded as the lowest energy along the dynamical path that includes all dynamical effects through the time evolution of the realistic TDHF density distributions.
In the DD-TDHF method, 
the potential energy is obtained under the assumption that  
complex TDHF dynamics for central collisions is reduced to one-dimensional macroscopic equation of motion. 
Effects coming from other collective degrees of freedom than the degree of freedom of the relative motion 
are reduced to the transport coefficients of the reduced mass, nucleus--nucleus potential, and dissipation term in the one-dimensional Newton equation.
Although both methods are based on the same TDHF dynamics,
the way of interpretation of nucleus--nucleus potential is totally different from each other, giving different properties of the potential in heavy systems.
Note that the reduced mass extracted from the DD-TDHF method includes dynamical effects, leading to the $R$ dependence of mass.

However, in light systems, we do not see a significant difference between DC-TDHF and DD-TDHF potentials. As an example, 
we show in Fig.~\ref{fig:potCa40Ca40compare} the comparison between DC-TDHF (blue dashed line) and our potential (red solid line) for the $^{40}$Ca\,+\,$^{40}$Ca system at $E_{\rm c.m.}=55$\,MeV, 
and find no significant difference between the two potentials 
(only 0.3\,MeV difference at most at the barrier).
Therefore, the effects mentioned above are significant only in heavy systems.

\subsection{Dissipation}\label{sec:result:dissipation}

The TDHF theory includes one-body dissipation through the selfconsistent mean field.
As we explained in Sec.~\ref{sec:method:DD-TDHF}, 
we can simultaneously extract the nucleus--nucleus potential and the friction coefficient from the DD-TDHF method.
In this subsection, we discuss the property of extracted friction 
coefficients in heavy systems. 

\begin{figure}[tb]
\includegraphics[width=0.9\linewidth,clip]{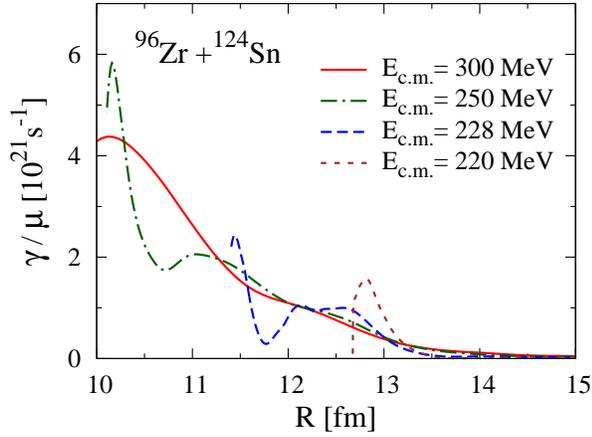}
\caption{\label{fig:frictionZr96Sn124} (Color online)
Reduced friction coefficient defined as $\gamma(R)/\mu(R)$ as a function of $R$
extracted from TDHF trajectories at $E_{\rm c.m.}=300$\,MeV (red solid line), $E_{\rm c.m.}=250$\,MeV (green dot-dashed line), $E_{\rm c.m.}=228$\,MeV (blue dashed line), and $E_{\rm c.m.}=220$\,MeV (brown dotted line) for the $^{96}$Zr\,+\,$^{124}$Sn system.
}
\end{figure}
\begin{figure}[tb]
\includegraphics[width=0.9\linewidth,clip]{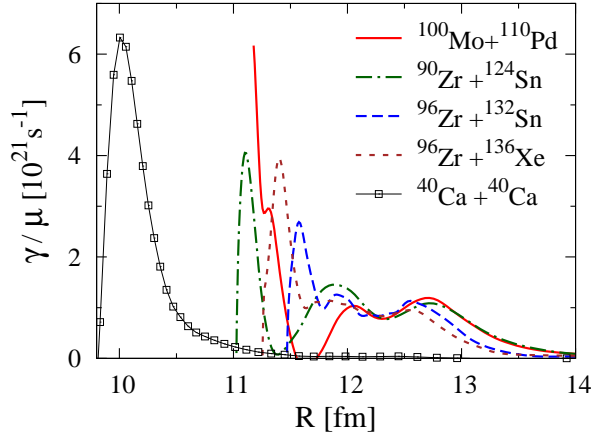}
\caption{\label{fig:friction} (Color online)
Reduced friction coefficient $\gamma(R)/\mu(R)$ as a function of $R$
for selected systems.
}
\end{figure}

In Fig.~\ref{fig:frictionZr96Sn124}, we show the reduced friction coefficients defined as the friction coefficient divided by the reduced mass $\gamma(R)/\mu(R)$ as a function of relative distance in the $^{96}$Zr\,+\,$^{124}$Sn system extracted at the same energies as used in extracting potential in Fig.~\ref{fig:potZr96Sn124}.
The extracted friction coefficient increases as the colliding nuclei approach at $R>13$\,fm.
And then, unlike the extracted potential,
we observe a clear energy dependence of the friction coefficient. 
As $E_{\rm c.m.}$ increases, the peak position of the extracted friction coefficient is shifted towards small $R$, which is close to $R_{\rm min}$, the relative distance at which we stop the extraction, at each energy.
(Note again that, for the cases at $E_{\rm c.m.}=250$, 228, and 220\,MeV, $R_{\rm min}=10.1$, 11.4, and 12.7\,fm, respectively.)
The reason of the appearance of the peak near $R_{\rm min}$ is that,
as $R$ becomes small,  the relative velocity becomes small and 
the internal structure of the colliding nuclei has much time to reorganize,
giving a large value of the friction coefficient.

Figure~\ref{fig:friction} shows the reduced friction coefficients for selected systems. 
For these systems, each friction coefficient is extracted at each fusion threshold energy. 
The radial dependence of the friction coefficients is similar in these systems:
As $R$ decreases, the friction coefficient increases,
makes two humps at $R\sim12.6$ and 12\,fm, and then has a sharp peak. 
Also, the values of their strengths at the humps are similar to each other, $\gamma/\mu \sim 1\times 10^{21}$\,s$^{-1}$,
and their strengths at the sharp peak are of the same order of magnitude, $\gamma/\mu \sim 2 - 6\times 10^{21}$\,s$^{-1}$.
For comparison, we also show the reduced friction coefficient for the 
$^{40}$Ca\,+\,$^{40}$Ca system at $E_{\rm c.m.}=55$\,MeV by the open squares.
The radial dependence of the friction coefficient is slightly different
from those in heavy systems: It has only a single peak.
Its strength is however of the same order of magnitude as that at the peak in heavy systems.
\begin{figure}[tb]
\includegraphics[width=0.9\linewidth,clip]{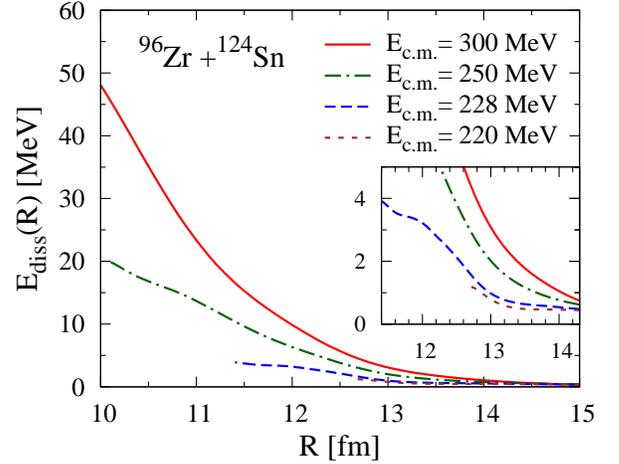}
\caption{\label{fig:Ediss} (Color online)
Accumulated dissipation energy calculated by Eq.~(\ref{eq:Ediss}) as a function of $R$ obtained from TDHF trajectories at $E_{\rm c.m.}=300$, 250, 228, and 220\,MeV for the $^{96}$Zr\,+\,$^{124}$Sn system. A zoom around $R=13$\,fm is inserted.
}
\end{figure}

With the extracted friction coefficient, we can evaluate accumulated dissipation energy from the relative motion to internal excitations by the formula:
\begin{equation}\label{eq:Ediss}
E_{\rm diss}[R(t)] = \int_0^t dt^\prime \gamma[R(t^\prime)] {\dot R}(t^\prime)^2,
\end{equation}
until time $t$ corresponding to the relative distance $R(t)$ in the entrance channel. 
Figure~\ref{fig:Ediss} shows the accumulated dissipation energy as a function of $R$ for 
the $^{96}$Zr\,+\,$^{124}$Sn system at the same $E_{\rm c.m.}$ as in Fig.~\ref{fig:frictionZr96Sn124}.  
A zoom around $R=13$\,fm is inserted in the figure to focus on a low $E_{\rm diss}$ region.
As expected from the property of the extracted friction coefficient, the dissipation energies are zero at large $R$ and then monotonically increase as the colliding nuclei approach and overlap each other.
Irrespective of humps and peaks of the extracted friction coefficients,
the dissipated energies show a smoothed behavior.
Concerning the dependence of dissipation energy on $E_{\rm c.m.}$ of TDHF trajectory, 
the dissipation energy becomes larger at larger $E_{\rm c.m.}$
because of larger relative velocity entering in the integral in Eq.~(\ref{eq:Ediss}). 

In the following, we will use the dissipation energy $E_{\rm diss}$ accumulated until $R_{\rm min}$ as the physical quantity to analyze the fusion hindrance in the entrance channel of fusion reactions.

\subsection{Origin of extra-push energy}
\begin{figure}[tb]
\includegraphics[width=0.8\linewidth,clip]{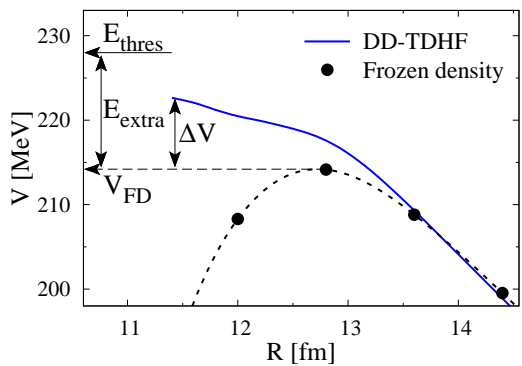}
\caption{\label{fig:pot.schematic} (Color online)
Schematic picture to define the quantities used in the analysis: Frozen density potential barrier $V_{\rm FD}$, fusion threshold energy $E_{\rm thres}$, extra-push energy $E_{\rm extra}=E_{\rm thres}-V_{\rm FD}$, and increase in potential $\Delta V$ for the $^{96}$Zr\,+\,$^{124}$Sn system.
}
\end{figure}

\begin{figure*}[tb]
\includegraphics[width=0.9\linewidth,clip]{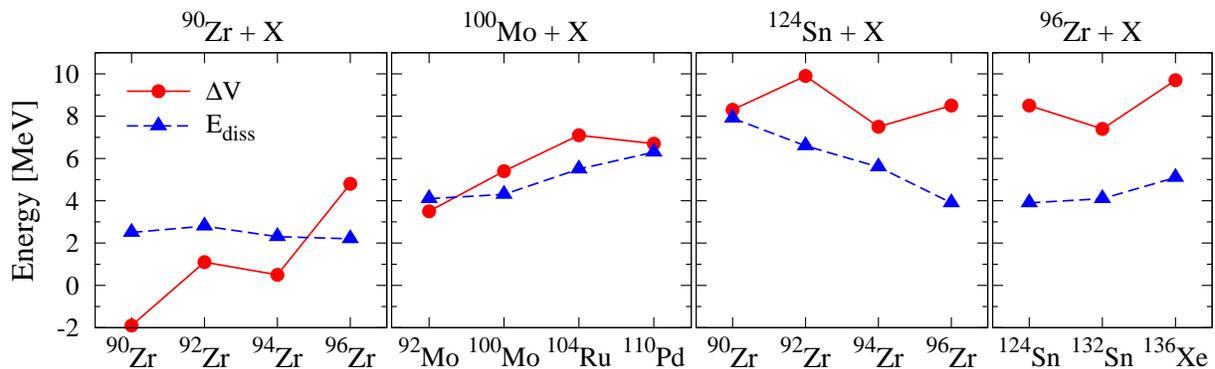}
\caption{\label{fig:dv.ediss}  (Color online)
Contributions of the increase in potential $\Delta V$ and of the accumulated dissipation energy $E_{\rm diss}$ to the extra-push energy for the same systems as those in Fig.~\ref{fig:extrapush}.
}
\end{figure*}

In Sec.~\ref{sec:result:extrapush}, we have observed the fusion hindrance phenomenon by TDHF simulations for the entrance channel of central collisions in heavy systems. 
We have shown in Fig.~\ref{fig:extrapush} the extra-push energies estimated from TDHF simulations for several heavy systems 
and reasonably reproduced the extra-push energies deduced from experimental observations of evaporation-residue cross sections.
In this subsection, using the properties of the extracted potentials and friction coefficients in heavy systems presented in the previous subsections, 
we will analyze the origin of the fusion hindrance and appearance of the extra-push energy.

Before showing the result of the analysis, 
we first summarize the quantities in a schematic picture in Fig.~\ref{fig:pot.schematic} that will be used in the analysis. 
They are the fusion threshold energy $E_{\rm thres}$ and the frozen density potential barrier $V_{\rm FD}$, indicated by the arrows, for the $^{96}$Zr\,+\,$^{124}$Sn system. We have defined the extra-push energy by TDHF as the difference between the fusion threshold energy and the frozen density potential barrier, $E_{\rm extra}=E_{\rm thres}-V_{\rm FD}$.
Here, we introduce a new quantity, increase in potential energy $\Delta V$, that is the difference between the extracted potential at $E_{\rm thres}$ at $R_{\rm min}$ and the frozen density potential barrier. 
Since the extracted potential increases monotonically as $R$ decreases in the heavy systems, $\Delta V$ becomes maximum at $R_{\rm min}$.
We will also use in the analysis the accumulated dissipation energy $E_{\rm diss}$ defined in Eq.~(\ref{eq:Ediss}).
This accumulated dissipation energy is computed from the friction coefficient 
extracted at $E_{\rm thres}$. As shown in Fig.~\ref{fig:Ediss}, $E_{\rm diss}$ also monotonically increases as $R$ decreases.

From the energy conservation, we have at $R\geq R_{\rm min}$ 
\begin{equation}\label{eq:energy_conservation}
E_{\rm c.m.} = E_{\rm kin}(R) + V(R) + E_{\rm diss}(R),
\end{equation}
where $E_{\rm kin}=P^2/2\mu(R)$ is the collective kinetic energy 
and all the energies in the right-hand side can be computed by TDHF and by the DD-TDHF method and we confirm that the energy conservation holds.
Subtracting $V_{\rm FD}$ from the both sides of Eq.~(\ref{eq:energy_conservation}) and using the quantities $E_{\rm extra}= E_{\rm thres}-V_{\rm FD}$ and $\Delta V=V(R_{\rm min})-V_{\rm FD}$,
we derive the following relation at $E_{\rm c.m.}=E_{\rm thres}$ at $R=R_{\rm min}$:
\begin{equation}
E_{\rm extra}=E_{\rm kin}(R_{\rm min})+ \Delta V(R_{\rm min})+ E_{\rm diss}(R_{\rm min})
\end{equation}
As shown in Fig.~\ref{fig:momentum}, the absolute value of the momentum $|P|$ is expected to monotonically decrease and to be smallest at $R=R_{\rm min}$ in heavy systems, 
so is $E_{\rm kin}$.
We have checked that $E_{\rm kin}$ at $R=R_{\rm min}$ in all the systems is less than 0.6\,MeV and small enough, compared with $\Delta V$ and $E_{\rm diss}$.
Therefore, we approximate the extra-push energy as the sum
of $\Delta V$ and $E_{\rm diss}$,
\begin{equation}
E_{\rm extra}\approx\Delta V(R_{\rm min})+ E_{\rm diss}(R_{\rm min}).
\end{equation}
That is, we can identify the origin of the extra-push energy in the entrance channel as $\Delta V$ and/or $E_{\rm diss}$ extracted from the DD-TDHF method.

Figure~\ref{fig:dv.ediss} shows each contribution to the extra-push energy
for the same systems as in Fig.~\ref{fig:extrapush}.
The red filled circles denote the contribution of the increase in potential $\Delta V$ to the extra-push energy, while the blue filled triangles denote the contribution of the accumulated dissipation energy $E_{\rm diss}$.
In the $^{90}$Zr\,+\,$^{90,92,94}$Zr and $^{100}$Mo\,+\,$^{92}$Mo systems, $\Delta V$ is smaller than $E_{\rm diss}$. 
In the rest of the systems considered here, $\Delta V$ is greater than $E_{\rm diss}$. 
As the charge product of the system increases, the ratio of $\Delta V$ to $E_{\rm diss}$ increases.
From this figure, we can conclude that, as the system becomes heavier, 
the contribution of the increase in potential energy 
to the extra-push energy in the entrance channel
is more important than that of the accumulated dissipation energy.
Dissipation plays an important role, but is not sufficient to explain the 
amount of the extra-push energy in our analysis with the DD-TDHF method.

\section{Conclusions}\label{sec:conclusions}

The synthesis of superheavy elements in laboratories is extremely challenging 
because of an extremely low production rate. One of the reason for a low production late
is the presence of the fusion hindrance. This gives strong competition
between compound nucleus formation and quasifission.
In this article we address this point by the selfconsistent microscopic reaction theory, TDHF.
We have shown the applicability of TDHF dynamics for fusion reactions involving heavy nuclei by reproducing the extra-push energies estimated from experimental evaporation-residue cross sections in heavy systems.
Then, we have analyzed the property of the nucleus--nucleus potentials for the entrance channel of fusion reactions
extracted from the DD-TDHF method, which combines TDHF dynamics with a classical equation of motion.
In heavy systems, the extracted nucleus--nucleus potentials show the following properties:
(i) Ordinary Coulomb barrier in the extracted potential disappears and the potential monotonically increases as the relative distance decreases.
(ii) The dependence of potential on the energy used in the TDHF simulations is not pronounced.
These properties are not seen in the potentials in light- and medium-mass systems and also in the DC-TDHF method.
We have shown that the increase in potential is a reflection of the appearance of the conditional saddle inside the contact configuration of colliding nuclei on an adiabatic potential surface.
By analyzing the isotope dependence of the extracted potentials,
we have found that, as nucleon transfers towards charge equilibration during the collision becomes small, increase in the extracted potential from the frozen density potential becomes large.
Extracted friction coefficients however show energy dependence, and have a sharp peak near the relative distance at which the extraction stops. This indicates strong reorganization of the internal structure near that distance.
Using these properties of the extracted nucleus--nucleus potential and energy dissipation, we have analyzed the origin of the extra-push energy in heavy systems.
The analysis has shown that more contribution comes from the increase in potential energy to extra-push energy than that from the accumulated dissipation energy
as the size of the system is heavier.
We have concluded that the increase in potential is more important than dissipation for understanding the origin of the fusion hindrance.

Although the present study is limited to central collisions, 
which enable us to perform the DD-TDHF method with one-dimensional equation of motion,  
we have shown the feasibility of the present analysis for the fusion hindrance.
Possible extensions to non-central collisions involving finite angular momenta will be interesting to understand the role of angular momentum for the fusion hindrance.
Furthermore, the orientation dependence of the dynamics due to large nuclear deformation will be important for systems with, for example, actinide nuclei, which have been and will be used in the so-called hot fusion reactions in the synthesis of superheavy elements.
The fission process, as an inverted process of fusion,
is one of the future subject to apply the present method to extract 
information on energy dissipation.

\begin{acknowledgments}
The author thanks D. Lacroix, T. Nakatsukasa, and K. Sekizawa for providing valuable comments on this article.
The author is supported by the Special Postdoctoral Researcher Program of RIKEN.
\end{acknowledgments}

\appendix

\section{Criterion of determining the neck position}\label{sec:appendix:neck}

To construct two-body kinematics in the DD-TDHF method in Sec.~\ref{sec:method:DD-TDHF}, 
we need to separate the total system into two subsystems at the neck position. 
In the case of central collisions along $x$-axis, 
the neck position can be specified by $x=x_0$.
Previously, we have determined the neck position
where the densities of the projectile and target cross 
and checked that this definition is good for light- and medium-mass systems at any energies \cite{washiyama08,washiyama09}, hereafter called criterion (i).
However, we realize that this criterion is inappropriate 
for some cases, especially at small relative distances in heavy systems.
Then, we test the following criteria to better determine the neck position:
(ii) The position where the total one-dimensional density $\rho(x,y=y_0,0)$ becomes minimum at $y_0=0$\,fm.
(iii) The same as criterion (ii) but at $y_0=2$\,fm.
By looking at the two-dimensional density profiles, we find that the criterion (iii) is most reasonable and stable for the cases at low energies in most heavy systems considered in this article.
Therefore, we use the criterion (iii) to determine the neck position
at low energies. For higher energies, we find that the criterion (i) is better. 

\section{Ambiguity of determining the fusion threshold energy}
\label{sec:appendix:threshold}

We have explained how to determine the fusion threshold energy in Sec.~\ref{sec:method:definition}. 
When we determine the fusion threshold energy,
we have encountered an ambiguity to define a compact shape in some systems.
In the following, 
we will show that the conclusion of our study, i.e., the increase in potential is more important than the accumulated dissipation energy for understanding the origin of the fusion hindrance, is not be affected by this ambiguity.
To do so, we take the $^{90}$Zr\,+\,$^{124}$Sn system as an example.
In this system, we indeed define the fusion threshold energy as $E_{\rm c.m.}=234.5$\,MeV.
The reason is that $Q_{20}$ at $E_{\rm c.m.}=234$\,MeV is 15\% larger than $Q_{20}$ at higher energies (see Fig.~\ref{fig:q20Zr90Sn124})
and we regard the configuration at $E_{\rm c.m.}=234$\,MeV as not a compact shape. 
Therefore, we exclude the case at $E_{\rm c.m.}=234$\,MeV from the fusion events.
However, one might consider this case as still a compact shape, if one compares this with the case at $E_{\rm c.m.}=233.5$\,MeV, in which the system reseparates.
In order to check how this ambiguity affects the result, 
we change the fusion threshold energy from 234.5\,MeV to 234\,MeV in this system and repeat the same calculation. 
We find that the increase in potential and accumulated dissipation energies are changed from 8.3 and 7.9\,MeV for $E_{\rm thres}=234.5$\,MeV to 8.0 and 7.7\,MeV for $E_{\rm thres}=234$\,MeV, respectively. 
The ratio of $\Delta V/E_{\rm diss}$ does not change significantly with the change of $E_{\rm thres}$ and we therefore hold the conclusion as it is.

\bibliography{/home/washi/Dropbox/reference}
\end{document}